%% file: main.tex
\documentclass{article}

\usepackage{arxiv}

\usepackage[utf8]{inputenc}
\usepackage[T1]{fontenc}
\usepackage{hyperref}
\usepackage{url}
\usepackage{booktabs}
\usepackage{amsmath,amsfonts}
\usepackage{amssymb}
\usepackage{algorithmic}
\usepackage{algorithm}
\usepackage{array}
\usepackage{multirow}
\usepackage{nameref}
\usepackage{mathtools}
\usepackage{graphicx}
\usepackage[numbers,sort&compress]{natbib}
\usepackage{doi}
\usepackage{microtype}
\usepackage{cleveref}

\input{notation}

\newlength{\colwidth}
\newcommand{\berDisclaimer}{
When an image is cropped, the payload in the cropped region is lost, making BER ill-defined. Therefore, we do not report BER in this cases.
}

\hypersetup{
  pdftitle={DeepSignature: Digitally Signed, Content-Encoding Watermarks for Robust and Transparent Image Authentication},
  pdfsubject={cs.CV, cs.CR},
  pdfauthor={Mathias Graf, Marco Willi, Melanie Mathys, Michael Aerni, Christian Schwarzer, Martin Melchior, Michael H Graber},
  pdfkeywords={Artificial intelligence, Deep learning, Watermarking, Digital signatures, Image authentication},
}

\renewcommand{\shorttitle}{DeepSignature}

\title{DeepSignature: Digitally Signed, Content-Encoding Watermarks for Robust and Transparent Image Authentication}

\date{April 2026}

\author{
  Mathias Graf\thanks{Co-first authors.} \\
  Institute for Data Science, FHNW \\
  \And
  Marco Willi\footnotemark[1] \\
  Institute for Data Science, FHNW \\
  \And
  Melanie Mathys \\
  Institute for Data Science, FHNW \\
  \And
  Michael Aerni \\
  Institute for Data Science, FHNW \& ETH Zürich \\
  \AND
  Christian Schwarzer \\
  Institute for Data Science, FHNW \\
  \And
  Martin Melchior \\
  Institute for Data Science, FHNW \\
  \And
  Michael H Graber\thanks{Corresponding author: \texttt{michael.graber@fhnw.ch}} \\
  Institute for Data Science, FHNW \\
}

\begin{document}
\maketitle

\begin{abstract}
    AI-powered generative models have significantly expanded the
    possibilities for editing, manipulating, and creating high-quality
    images. Particularly, images that falsely appear to originate from trusted sources pose a serious threat, undermining public trust in image authenticity.
    We propose DeepSignature, a novel approach that integrates the guarantees of digital signatures with the capabilities of deep neural networks.
    Neural networks are used both to generate content-encoding watermarks and to embed them imperceptibly into images while ensuring robust extraction. These watermarks are cryptographically verifiable, enabling source attribution and image integrity validation.
    DeepSignature is compatible with existing image formats and requires no special handling of signed images. It supports client-side verification, requiring only the signer's public key. Additionally, we introduce a novel latent-space verification approach to detect and localize tampering attempts.
    We evaluate DeepSignature in terms of imperceptibility, robustness to benign transformations, forgery detection, and its resilience against various attack scenarios. Our results highlight the inherent trade-offs between imperceptibility, robustness, and integrity verification. We demonstrate that DeepSignature reliably identifies significant forgery attempts---achieving near 100\% in our experiments. Finally, we emphasize DeepSignature’s modularity and tunable parameters, allowing adaptation to application-specific requirements.
    Code and model weights will be published.
\end{abstract}


\section{Introduction}

\begin{figure}[!t]
    \centering
    \includegraphics[width=\linewidth]{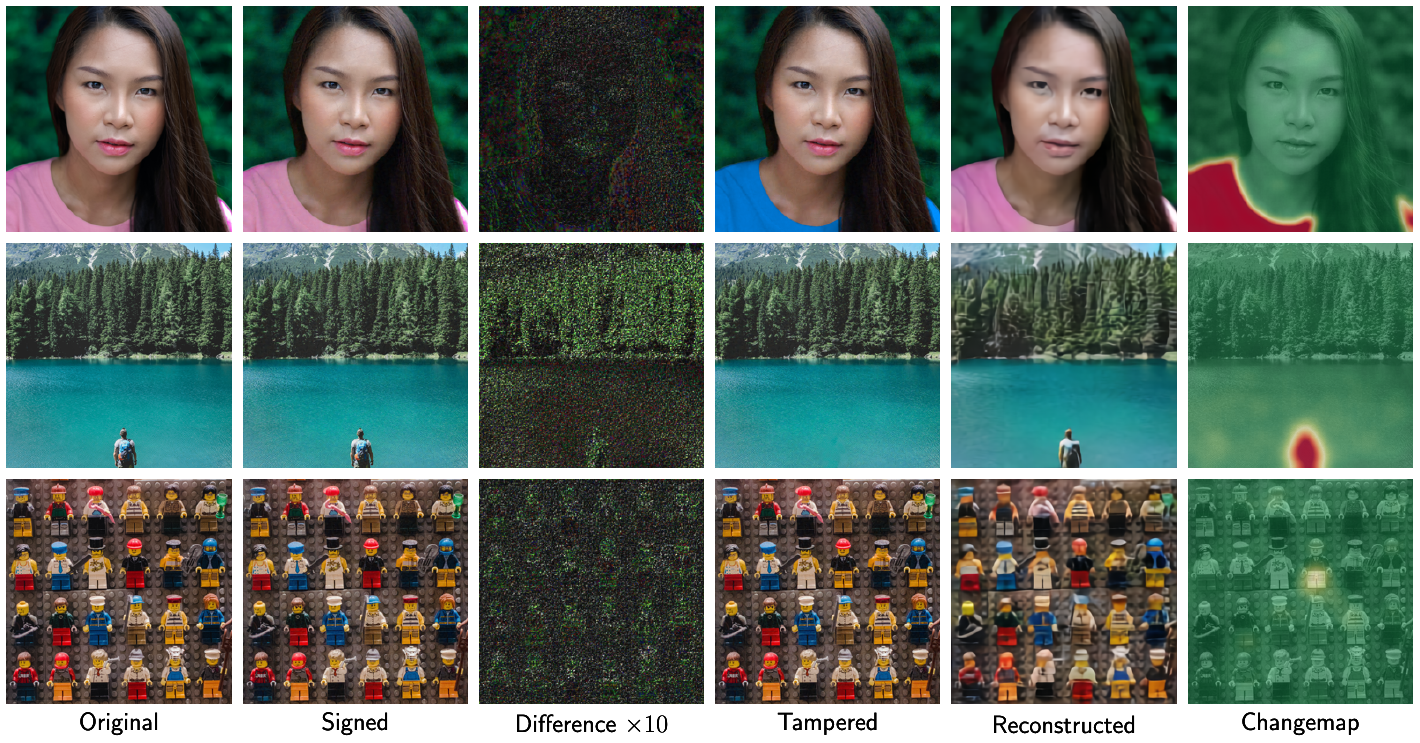}
    \caption{DeepSignature reliably detects and localizes tampering in signed images. The figure presents results for three images, each shown in a separate row. From left to right, each row displays: the original image (\textit{Original}), the DeepSignature-signed image (\textit{Signed}), the magnified absolute difference between the original and signed images (\textit{Difference $\times 10$}), a tampered version of the signed image (\textit{Tampered}), the reconstructed original image obtained from the authenticated content-encoding watermark hidden in the tampered image (\textit{Reconstructed}), and a change map overlaid on the tampered image, highlighting localized tampering scores  (\textit{Changemap}). All images are $768 \times 768$ pixels.
    Created using images from CLIC 2022 and CLIC 2024 test sets licensed under the Unsplash License.
    }
    \label{fig:examples}
\end{figure}

Tremendous progress in AI-powered generative models has supercharged the possibilities of editing, manipulating, and creating images of high quality and with minimal effort \citep{rombachHighResolutionImageSynthesis2022}. Generative models and their capabilities have entered public awareness, thereby degrading public trust in image authenticity. Providing the means to confirm source and integrity (together \textit{authenticity}) of images has thus become a pressing issue. Policymakers, academics, and industry stakeholders are actively working to protect image authenticity through initiatives like the Coalition for Content Provenance and Authenticity (C2PA) \citep{c2pa_2021}.

Digital signatures guarantee image authenticity using signed cryptographic hashes of the image content \citep{friedmanTrustworthyDigitalCamera1993}. To authenticate an image the signature needs to be available, for example, as attachable metadata. Practical limitations involve metadata stripping when images are processed or the invalidation of the signature upon any bit-level modification, such as image compression. Furthermore, if a signature can not be authenticated, there is no additional information beyond this binary decision.

Watermarking embeds the information necessary for image authentication directly into the image, often imperceptibly. A decoder analyses images for watermarks and extracts relevant information. Watermarks might encode a unique identifier to index a provenance database or they directly protect image integrity by enabling tampering detection and/or localization. Watermarking eliminates the need for additional metadata, ensuring that authentication information remains embedded even when images are re-encoded or stripped of metadata. However, watermarking requires careful balancing of imperceptibility, robustness to image transformations, and protection against adversarial attacks.

Our method, DeepSignature, combines the guarantees of cryptographic digital signatures, the convenient transport mode of watermarking, and the capabilities of deep neural networks to compress (also \textit{encode}) content. DeepSignature enables robust and transparent verification of image authenticity by essentially: hiding a compressed and signed version of the original image in itself.

Our technical contributions are the following:
\begin{enumerate}
    \item \textbf{Content-Encoding Watermarks}. We use a neural network to generate a content-encoding watermark, which we digitally sign and imperceptibly encode into a target image using a neural network encoder. The signature provides cryptographic guarantees with respect to source and integrity of the watermark.

    \item \textbf{Latent-space verification}. We propose a latent-space image integrity verification method to compare the received (and potentially manipulated) image with the content encoded in the watermark. We demonstrate that we reliably detect and localize forgery attempts.

    \item \textbf{Application specific trade-offs}: We propose mechanisms to enable trade-offs between imperceptibility, robustness, and verifiability to tailor DeepSignature to application-specific requirements.
\end{enumerate}

Figure~\ref{fig:examples} depicts several example images, including their signed versions, tampering attempts, and how DeepSignature reveals and localizes the manipulations.

\section{Related Work}

\subsection{Image Authenticity Verification}

Verifying image authenticity, i.e. enabling source attribution and integrity validation, is crucial in many applications. Broadly, two approaches address this challenge: passive and active methods \citep{korusDigitalImageIntegrity2017}.

Passive (forensic) methods analyze images for synthetic traces \citep{chenDeterminingImageOrigin2008, guillaroTruForLeveragingAllRound2023, wangCNNgeneratedImagesAre2020}. They can be applied to any image, however, require constant adaptation to novel generative models \citep{epsteinOnlineDetectionAIGenerated2023} and are susceptible to counter-forensics methods \citep{carliniEvadingDeepfakeImageDetectors2020}. Ultimately, there is no guarantee for their accuracy.

Active methods protect digital media through digital signatures or watermarks \citep{coxDigitalWatermarkingSteganography2008}. Unlike passive techniques, they can provide verifiable authenticity guarantees or pointers to provenance databases, but only for images that were \textit{actively} protected. Guidelines by the C2PA \citep{c2pa_2021}, for example, outline how to secure digital assets with digital signatures.

\subsection{Steganography and Watermarking}
Image steganography aims to hide secret messages in images, there\-by balancing trade-offs among imperceptibility, message length, and secrecy---that is, minimizing the detectability of images with hidden messages. Techniques such as the least significant bit (LSB) method have historically been used to embed secret messages in the lowest-order bits of encoded pixel values \citep{chanHidingDataImages2004}. These methods, however, are susceptible to detection \citep{qinReviewDetectionLSB2010} and lack robustness against image transformations.

Watermarking for image authentication is concerned with imperceptibly embedding secret messages or signals into digital assets. Watermarks may be designed to withstand image transformations (robust watermarks), to destroy under any transformations (fragile watermarks), or to be robust to selected transformations (semi-fragile watermarks). They can also enable tampering detection or localization (authentication watermarking), or even recover the original appearance (self-recovery watermarking) \citep{korusDigitalImageIntegrity2017}.

\subsection{Data Hiding With Deep Learning}
Deep learning has had a profound impact on both steganography and watermarking approaches. For example, HiDDeN \citep{zhuHiDDeNHidingData2018} demonstrated that end-to-end learned models achieve superior secrecy for steganography, as well as strong robustness to selected transformations by exposing models to differentiable transformations during training. Furthermore, these models retain high perceptual fidelity by using perceptual and/or adversarial loss functions. Some approaches have demonstrated remarkable capabilities, such as hiding entire images within cover images \citep{balujaHidingImagesPlain2017}. Recent works concerned with watermarking and image authenticity verification improved different aspects of data hiding, such as improved robustness and/or imperceptibility~\citep{luoDistortionAgnosticDeep2020, tancikStegaStampInvisibleHyperlinks2020, buiTrustMarkUniversalWatermarking2023, jiaMBRSEnhancingRobustness2021, xuInvisMarkInvisibleRobust2024}, novel ways to leverage pre-trained models~\citep{fernandezWatermarkingImagesSelfSupervised2022, buiRoSteALSRobustSteganography2023}, or hiding semantic information (text captions)~\citep{evennouSWIFTSemanticWatermarking2024} to detect tampering.

Data hiding methods involve trade-offs among hiding capacity, imperceptibility, and robustness. Increasing payload capacity and robustness typically reduces imperceptibility; therefore, application-specific operations points must be chosen.
To compare methods with different message lengths and bit error rates (BER), we approximate mean BER under a specific transformation as the crossover probability $p$ of a binary symmetric channel (BSC), and compute an effective achievable payload size (in bits per pixel) as \citep{tancikStegaStampInvisibleHyperlinks2020}:

\begin{equation}
    C_\mathrm{BSC} = \frac{m}{H \times W}\,\bigl(1 - H_b(p)\bigr),
\end{equation}

where $m$ is the message length, $H$ and $W$ are width and height of the image, $p$ the crossover probability (BER) and $H_b$ is the binary entropy function:

\begin{equation}
    H_b(p) = -p  \log_2(p) - (1 - p) \log_2(1 - p).
\end{equation}

Recent works have demonstrated watermarking with capacities in the range from 30 -- 4000 bits per megapixel under JPEG quality 50 \citep{buiRoSteALSRobustSteganography2023, jiaMBRSEnhancingRobustness2021, evennouSWIFTSemanticWatermarking2024, fernandezWatermarkingImagesSelfSupervised2022, xuInvisMarkInvisibleRobust2024}. Such capacities are sufficient to embed a unique identifier but not content-specific information. Naturally, methods for low-capacity watermarking prioritize achieving both high robustness and low perceptibility.

\subsection{Image Encoding}
Image encoding describes the process of transforming an image into a latent representation. Images (i.e., photographs) are often high dimensional, with strong (local) correlations and high-frequency details. This characteristic enables effective perceptual compression that preserves an image's semantic content \citep{rombachHighResolutionImageSynthesis2022}. Various domains---such as neural image compression~\citep{mentzerHighfidelityGenerativeImage2020}, image retrieval~\citep{radfordLearningTransferableVisual2021}, perceptual hashing~\citep{struppekLearningBreakDeep2022}, or generative modeling~\citep{rombachHighResolutionImageSynthesis2022}--exploit perceptual compression to varying degrees.
Latent representations, which capture the essential features of an image while discarding semantically irrelevant details, are also employed in the self-supervised pre-training of vision models~\citep{mahajanExploringLimitsWeakly2018}. This approach enables high performance in downstream tasks.

There are various methods to learn latent representations, including autoencoders~\citep{kingmaAutoEncodingVariationalBayes2014} and contrastive modeling~\citep{chenSimpleFrameworkContrastive2020}. Autoencoders learn to encode images by jointly optimizing both an encoder and a decoder to reconstruct the input image. Recent advances include vector quantized autoencoders~\citep{vandenoordNeuralDiscreteRepresentation2017}, which learn discrete latent representations, thereby enabling straight-forward conversion to compact binary representations. The compression rate (i.e., the bit-ratio between the input and latent size) can be precisely controlled and tailored to a specific purpose.

\section{Method}

\begin{figure*}[!t]
    \centering
    \includegraphics[width=\textwidth]{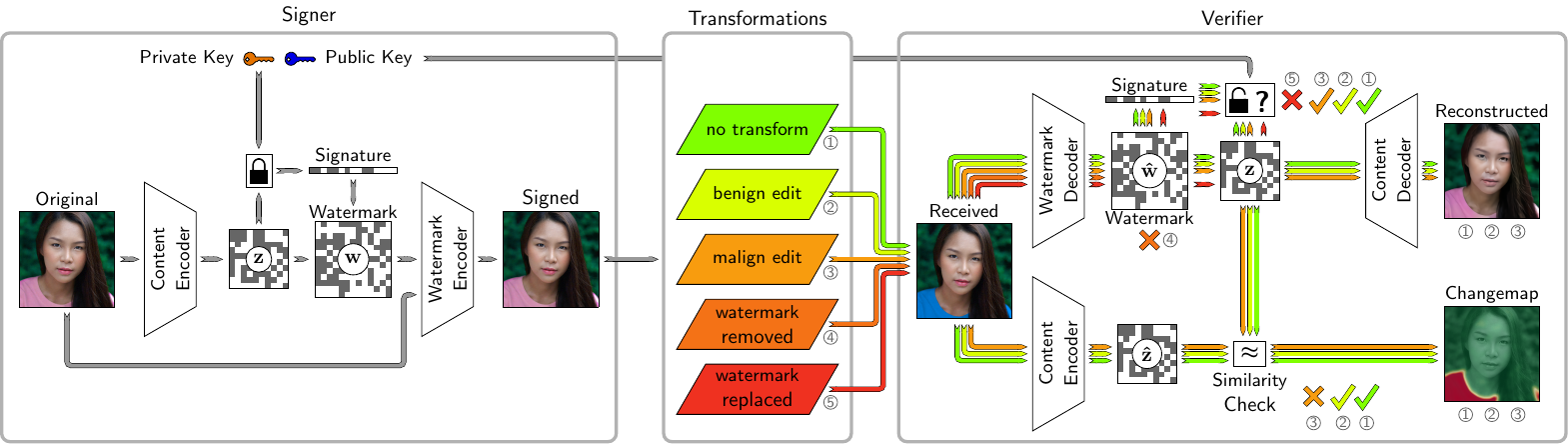}
    \caption{Illustration of DeepSignature. The left panel illustrates the signer's workflow, which involves encoding an original image into a latent representation ($\mathbf{z}$), digitally signing $\mathbf{z}$ using the signer's private key, and embedding both the signature and $\mathbf{z}$ as a watermark ($\mathbf{w}$) into the image. This process results in a signed image. The middle panel depicts how the signed image may undergo various transformations or attacks (represented by colored boxes and pathways). The right panel presents the verifier's workflow. The verifier receives an image, extracts the watermark ($\hat{\mathbf{w}}$), and verifies it cryptographically using the signer's public key. The watermark is then used to assess the integrity of the received image by comparing its latent representation ($\hat{\mathbf{z}}$) with the original one ($\mathbf{z}$). The encoder and decoder components are implemented as neural networks.
    For images without transformations or with benign edits both the cryptographic verification of the watermark and the image similarity check succeed, for images with malign edits the watermark verification succeeds but the similarity check fails and in cases of removed or replaced watermarks the cryptographic verification of the watermark fails.
    Created using an image from CLIC 2024 test set licensed under the Unsplash License.
    }
    \label{fig:teaser}
\end{figure*}

\noindent DeepSignature uses an autoencoder for image encoding, a watermark encoder to embed the signed encoding into the image, a watermark decoder to retrieve the encoding, and the image encoder to compare the received, and potentially tampered with, image with the signed encoding. The overall architecture is depicted in Figure~\ref{fig:teaser}.

\subsection{Content Encoding / Decoding}

We use a vector quantized variational autoencoder (VQ-VAE) \citep{vandenoordNeuralDiscreteRepresentation2017} to train a content encoder $\ContentEncoder$ which maps an image $\OriginalImageLong$ to a latent, continuous representation
 $\ContentEncoderOutputLong$. The content encoder is a convolutional neural network with global stride $\DownsamplingFactor$, thus $H' = \frac{H}{\DownsamplingFactor}$ and $W '= \frac{W}{\DownsamplingFactor}$. The $D$-dimensional vectors at each spatial location of $\ContentEncoderOutput$ are replaced with the index of the corresponding closest vector in the VQ-VAE's codebook $\CodebookLong$ to get the quantized representation $\QuantizedRepresentationLong$. The content decoder $\ContentDecoder$ is trained to reconstruct $\OriginalImage$ from $\ContentDecoderInputLong$, which is a latent, continuous but quantized representation using vectors from the codebook indexed by $\QuantizedRepresentation$. We refer to this quantization operation as $\Quantizer(\ContentEncoderOutput) = \ContentDecoderInput$. As a result, the output of the content encoder $\ContentEncoderOutput$, is approximately equal to $\ContentDecoderInput$:
 \begin{equation}
     \OriginalImage \approx \ContentDecoder\big(\Quantizer(\ContentEncoder(\OriginalImage))\big) \, .
 \end{equation}

 To reduce the spatial dimensionality of $\QuantizedRepresentationLong$, the input image $\OriginalImage$ may be scaled using bicubic interpolation. This can be achieved using a scaling factor $\ScalingFactorLong$, which reduces the size of the latent representation by $\ScalingFactor^2$.

\subsection{Digital Signature}

The quantized representation $\QuantizedRepresentationLong$ is first binarized to $\BinaryEncodingLong$, where $l = H'W'\log_2(\CodebookSize)$. The resulting binary encoding $\BinaryEncoding$ is then digitally signed using public-key cryptography to produce the signature $\Signature$. We employ the Edwards-curve Digital Signature Algorithm (EdDSA)~\citep{josefssonEdwardsCurveDigitalSignature2017} with pre-hashing (Ed25519ph), where $\BinaryEncoding$ is hashed using SHA512 and then signed using the private key $\PrivateKey$:
\begin{align}
\Signature &= \text{Sign}(\PrivateKey, \BinaryEncoding)  \, .
\end{align}

This approach ensures that the latent representation cannot be forged or modified by an attacker. We choose Curve25519 for EdDSA due to its efficient trade-off between security and signature length. It provides a 128-bit security level with a compact 512-bit signature, whereas an equivalent RSA-based PKCS1-v1\_5 signature would require 3072 bits~\citep{barkerRecommendationKeyManagement2016}. This makes EdDSA with pre-hashing an efficient choice for DeepSignature.

\subsection{Channel Coding, Metadata and Payload}
\label{sec:method:channel_coding}

A key goal of DeepSignature is to preserve the hidden watermark under various perturbations of the signed image. We view the process of encoding, transmitting, and decoding the watermark as communication over a noisy binary symmetric channel, where each bit has a probability $p$ of being \textit{flipped}. The watermark, however, remains verifiable only if \textit{every bit} can be correctly recovered. To achieve this, we apply forward error correction (channel coding), allowing watermark recovery even when $p>0$.

Technically, we concatenate the signature and the binarized content encoding to produce the watermark: $\Watermark = \BinaryEncoding \concat \Signature$, where $\WatermarkLong$. For channel coding, we use the Bose-Chaudhuri-Hocquenghem (BCH)~ \citep{boseClassErrorCorrecting1960} algorithm, because it is well established and implementations are readily available. We denote the channel coding function with $\ErrorCorrection_{\mathbf{c}}$, with algorithm-specific parameters $\mathbf{c}$, and it's inverse $\ErrorCorrectionDecoding_{\mathbf{c}}$. We enable image-specific, variable-length watermarks $\Watermark$ by specifying and hiding additional metadata $\MetadataLong$ which contains information about the exact length of $\Watermark$, as well as channel coding parameters $\mathbf{c}$. We use error correction $\ErrorCorrectionMetadata$ with image agnostic parameters for $\Metadata$. We denote the final bit-string that we hide in the image as the payload: $\Payload = \ErrorCorrectionMetadata(\Metadata) \concat \ErrorCorrectionWatermark(\Watermark)$, where $\PayloadLong$. Appendix~\ref{sec:payload_composition} contains additional details about how the payload $\Payload$ is constructed and decoded, and how some level of crop-resistance is achieved.

For simplicity, we refer to \textit{watermark} $\Watermark$ instead of payload $\Payload$ in the following sections, assuming channel coding and decoding, as well as de- and concatenation steps were successfully completed.

\subsection{Watermark Encoding / Decoding}
\label{sec:method:watermark_encoding_decoding}

We hide and extract data in and from images using the approach described in \citep{jiaMBRSEnhancingRobustness2021} (MBRS). We use a convolutional neural network with two components: a watermark encoder $\WatermarkEncoder$ and a watermark decoder $\WatermarkDecoder$. The encoder embeds a watermark---more precisely, the payload $\Payload$ (see Section \ref{sec:method:channel_coding})---into an image $\OriginalImage$, while the decoder extracts $\Payload$ from a watermarked image $\WatermarkedImage$. The watermarked image may have undergone noise transformations $\NoiseFunction$, such as JPEG compression. To ensure robustness and recoverability of $\Payload$ under benign transformations, we jointly train $\WatermarkEncoder$ and $\WatermarkDecoder$ with different $\NoiseFunction$:
\begin{equation}
    \Payload \approx \WatermarkDecoder\Big(\NoiseFunction\big(\WatermarkEncoder(\OriginalImage, \Payload)\big)\Big) \, .
\end{equation}

Note that the payload $\Payload$ does not need to be perfectly recovered by the decoder, since we tolerate some error through channel coding.

To provide some measure of control with respect to robustness, we enable scaling of the residual added to the original image by the watermark encoder by a strength factor $\StrengthFactor$. For $\StrengthFactor > 1$, robustness increases while perceptual quality decreases, and vice versa for $\StrengthFactor < 1$. This mechanism was introduced in \citep{jiaMBRSEnhancingRobustness2021}. A reasonable range might be $\StrengthFactor \in [0.8,1.5]$, as lower values lead to poor robustness and higher values to strong artifacts.

\subsection{Watermark Verification}
\label{sec:method:watermark_verification}

Given a (potentially) signed---i.e., watermarked---image, the verifier performs the following steps with the received image $\ReceivedImage$. The process begins with extracting the hidden watermark $\Watermark$. The verifier uses the watermark decoder $\WatermarkDecoder$ and attempts to extract $\Watermark$ (including decoding steps as eluded to in Section~\ref{sec:method:channel_coding}), and retrieves both the signature $\Signature$ and the latent representation $\BinaryEncoding$. Next, the receiver verifies the digital signature by using the  signer's public key $\PublicKey$ to check the integrity of $\BinaryEncoding$:
\begin{equation}
    \text{Verify}(\PublicKey,\BinaryEncoding,\Signature)\stackrel{?}{=}\text{True} \, .
\end{equation}

Watermark verification fails for any images i) without watermark, ii) large changes between $\WatermarkedImage$ and $\ReceivedImage$, iii) watermarked images subject to successful watermark removal attempts, and iv) watermarked images signed with a non-matching public key.

Successful watermark verification guarantees the authenticity of the hidden content encoding of the original image $\BinaryEncoding$, but does not ensure the integrity of the received image $\ReceivedImage$.

\subsection{Image Integrity Verification}
\label{sec:method:image_integrity}
To verify the integrity of the received image $\ReceivedImage$, it must be \textit{sufficiently} close to the original image $\OriginalImage$. Upon successful watermark verification (see Section~\ref{sec:method:watermark_verification}) the extracted and verified content encoding $\BinaryEncoding$ is converted to the quantized continuous representation $\ContentDecoderInput$. The verifier encodes the received image with the content encoder:
\begin{equation}
    \ContentDecoderInputReceived  = \ContentEncoder(\ReceivedImage) \, .
\end{equation}

The dimensionality of $\ContentDecoderInputReceived$ and $\ContentDecoderInput$ are identical\footnote{This is not the case if the received image $\ReceivedImage$ was cropped. In this case $\ReceivedImage$ is padded to its original size using the decoded metadata prior to encoding with $\ContentEncoder$}, which enables the calculation of a change map $\ChangeMapLong$. It represents the cosine dissimilarity at each spatial location of the received and the original image with respect to their encoded representation:
\begin{equation}
\label{eq:changemap}
    \ChangeMap^{(1, i,j)} = 1 - \left| \cos\left(\ContentDecoderInput^{(1, i,j)}, \ContentDecoderInputReceived^{(1, i,j)}\right)\right|  \, .
\end{equation}

The change map can be used to spatially pinpoint potential manipulations, for example, by rescaling and overlaying it onto the received image (see Figure~\ref{fig:examples}, last column). Additionally, either of the pairs $\ContentDecoder(\ContentDecoderInputReceived)$ / $\ContentDecoder(\ContentDecoderInput)$ or $\ReceivedImage$ / $\ContentDecoder(\ContentDecoderInput)$ can be compared visually in pixel-space (see examples in Figure~\ref{fig:examples}). Based on the change map $\ChangeMap$ we calculate a tampering score $\TamperingScoreLong$:
\begin{align}\label{eq:tamperingscore}
\ChangeMapAverage &= \frac{1}{H'W'} \sum_{\mathclap{i,j}} \ChangeMap^{(1,i,j)} \\
\TamperingScoreLocal &= \max(\ChangeMap) \\
\TamperingScoreGlobal &= \min(\TamperingScoreWeight \cdot \ChangeMapAverage, 1) \\
\TamperingScore &= \max(\TamperingScoreLocal, \TamperingScoreGlobal)
\end{align}

The $\TamperingScoreLocal$ is sensitive to local image manipulations
as seen in the examples from Figure~\ref{fig:examples}. $\TamperingScoreGlobal$ is relevant in case large image areas are affected by a manipulation of smaller magnitude. We chose $\TamperingScoreWeight = 3$ based on empirical tests on individual images, using JPEG as a proxy for benign changes and a small tampering dataset, similar to Figure~\ref{fig:examples}, for malign changes.

To verify image integrity, the tampering score $\TamperingScore$ can be compared to an application-specific threshold $\TamperingScoreThreshold$. Integrity verification of a watermarked image should fail for images with i) malign edits or ii) watermarks that do not match their content (e.g., watermarks extracted from other images, signed with $\PrivateKey$), and it should pass for images with iii) no transforms or iv) benign edits. We propose to select an optimal threshold according to the F1-score on a sample set.

\subsection{Loss Functions}
The content autoencoder, as well as the watermark models are trained separately, each optimized with their respective loss function. To provide an overview of the loss functions used, we detail them in this section.

\textbf{Content Encoder and Decoder:} We jointly train the content encoder $\ContentEncoder$ and decoder $\ContentDecoder$ to minimize the expectation over the following loss function:

\begin{align}
    \label{eq:loss_content_encoder_decoder}
    \begin{split}
    \mathcal{L}_{\ContentEncoder, \ContentDecoder} =
    \mathbb{E}_{\OriginalImage}
    \big[
    & \text{MSE}(\OriginalImage, \ReconstructedImage)
     + \text{SSIM}(\OriginalImage, \ReconstructedImage) \\
    & + \beta \cdot \text{MSE}(\ContentEncoderOutput, sg(\ContentDecoderInput)) \big]  \, .
    \end{split}
\end{align}

We penalize differences between the original image $\OriginalImage$ and the decoded version $\ReconstructedImage = \ContentDecoder\big(\Quantizer(\ContentEncoder(\OriginalImage))\big)$. MSE refers to the mean squared error and SSIM refers to the Structural Similarity Index Measure \citep{wangImageQualityAssessment2004}. The last term refers to the commitment loss from \citep{vandenoordNeuralDiscreteRepresentation2017}. We choose $\beta=0.25$.

\textbf{Watermark Encoder and Decoder:} We jointly train the watermark encoder $\WatermarkEncoder$ and decoder $\WatermarkDecoder$ to minimize the expectation over the following loss function:
\begin{align}
    \label{eq:loss_watermark_encoder_decoder}
    \begin{split}
    \mathcal{L}_{\WatermarkEncoder, \WatermarkDecoder} = \mathbb{E}_{\OriginalImage, \Watermark} \big[
    &\text{MSE}(\OriginalImage, \WatermarkedImage)
    + \beta_w \cdot \text{MSE}(\Watermark, \DecodedWatermark) \\
    & + \beta_a \cdot \log \big(1 - \AdversarialNetwork(\WatermarkedImage)\big)
    \big]  \, .
    \end{split}
\end{align}
We optimize for imperceptibility using MSE and a discriminator network $\AdversarialNetwork$ with an adversarial loss for the difference between the original image $\OriginalImage$ and the watermarked image $\WatermarkedImage$.
To achieve a low bit error rate (BER), we minimize MSE between the encoded watermark $\Watermark$ and the extracted watermark $\DecodedWatermarkLongBeforeQuant$. We set $\beta_a=10^{-3}$ and sample $\Watermark$ randomly from a Bernoulli distribution with $p=0.5$. During training, the weight $\beta_w$, which balances imperceptibility and $BER$, is adaptively adjusted using a proportional-integral (PI) controller to target a specific $BER_{target}$. The PI controller updates  $\beta_w$ based on the deviation $e(t) = BER(t) - BER_{target}$ at each iteration $t$.
It is configured using three parameters: the proportional gain $K_p$, which reacts to the current error; the integral gain $K_i$, which accumulates past errors to prevent steady-state bias; and the feedforward gain $K_f$, which scales the influence of $BER_{target}$ for direct adjustments.
Following \citep{jiaMBRSEnhancingRobustness2021}, we train the watermark encoder/decoder with transformations to enhance robustness. The noise function $\NoiseFunction$ randomly applies identity, differentiable JPEG approximation, or real JPEG compression. This improves resistance to JPEG and other distortions like compression artifacts, blurring, and contrast changes.

\section{Experiments and Results}

\noindent In this section we present experiments and results of different aspects of DeepSignature.

\subsection{Experimental Configuration}

For all experiments, unless stated otherwise, we use the following configuration and training setup for DeepSignature.

\textbf{Content Encoder/Decoder:} The content encoder $\ContentEncoder$ and decoder $\ContentDecoder$ are trained on the COCO 2017~\citep{linMicrosoftCOCOCommon2015} dataset by randomly cropping or padding the images to $512 \times 512$ pixels. We use a global stride of $S=16$, resulting in latent representations $\ContentEncoderOutput \in \mathbb{R}^{64 \times 32 \times 32}$ using a codebook $\CodebookLong$ with $K=256,D=64$. The latent representation thus has a fixed bitrate of $\frac{1}{32}$ bits per pixel. The training takes 29 hours on two Nvidia A4500 GPUs.

\textbf{Watermark Encoder/Decoder}: The watermark encoder $\WatermarkEncoder$ and decoder $\WatermarkDecoder$ are trained on 10,000 images from the COCO 2017 dataset, with each image randomly cropped to $128 \times 128$ pixels.
The encoder has a fixed capacity of $\frac{1}{(2^n)^2}$ bits per pixel, where $n$ represents the number of upsampling blocks in the watermark pre-processing sub-network. With $n=2$, the capacity is $\frac{1}{16}$ bits per pixel.
We use a 1024-bit watermark $\Watermark$, where each bit was randomly sampled from a Bernoulli distribution ($p=0.5$). The PI-controller for dynamic loss weighting is configured with $K_f=1$, $K_p = 100$, $K_i = 1$, and $BER_{target}=0.03$.
Training takes 6.5 hours on two Nvidia RTX 3080 GPUs.

The end-to-end process of signing one image of the CLIC 2024 test set in original size\footnote{2048 by 1388 pixels} on a Nvidia RTX3080 Laptop GPU takes around 3\,s, verification takes between 1.5\,s for an uncropped image and up to 12\,s for a cropped image. Cropped images can take significantly longer because the watermark decoder has to be run multiple times, refer to appendix~\ref{sec:cropresist} for details.

\subsection{Imperceptibility}

We measure Peak Signal-to-Noise Ratio (PSNR), Structural Similarity Index Measure (SSIM)~\citep{wangImageQualityAssessment2004}, its multi-scale version MS-SSIM~\citep{wangMultiscaleStructuralSimilarity2003}, and Learned Perceptual Image Patch Similarity (LPIPS)~\citep{zhangUnreasonableEffectivenessDeep2018} between the original image $\OriginalImage$ and its watermarked version $\WatermarkedImage$. These metrics quantify the perceptual degradation introduced by watermark encoding. We compare DeepSignature to JPEG image compression and evaluate different values for strength factor $\StrengthFactor$ (see Section~\ref{sec:method:watermark_encoding_decoding}).

\begin{figure}[!t]
    \centering
    \includegraphics[width=1\colwidth]{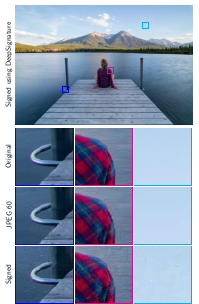}
    \caption{DeepSignature introduces more uniformly distributed noise artifacts compared to JPEG compression. The figure shows a DeepSignature-signed image, with three zoomed-in regions for each version: the original image (top row), a JPEG-compressed version (quality factor 60, middle row), and the DeepSignature-signed version (bottom row).
    Created using an image from the CLIC 2024 test set, licensed under the Unsplash License.
    }
    \label{fig:perceptibility_examples}
\end{figure}

Figure~\ref{fig:perceptibility_examples} shows one example image with zoomed-in regions to compare visual quality between the original image, a JPEG compressed version (quality factor 60), and a version signed with DeepSignature. Notice that JPEG introduces block artifacts while the artifacts from DeepSignature are spread more homogeneously.
Table~\ref{table:imperceptibility} shows the different metrics on the test set of the CLIC 2024 image compression challenge~\citep{clic_dataset_2024}.

\begin{table}[t]
  \caption{Imperceptibility measures. We report discrepancy between $\OriginalImage$ and $\WatermarkedImage$ using different metrics under variations of strength factor $\StrengthFactor$. $\uparrow$ (higher is better) and $\downarrow$ (lower is better). Results are reported on the CLIC 2024 test set.}
  \label{table:imperceptibility}
  \centering
  \small
  \setlength{\tabcolsep}{5pt}
\begin{tabular}{rrrr}
\toprule
 & PSNR$\uparrow$ & SSIM$\uparrow$ & LPIPS$\downarrow$ \\
 \midrule
Ours $\StrengthFactor$ = 0.8 & $36.75 \pm 2.01$ & $0.93 \pm 0.02$ & $0.10 \pm 0.05$ \\
Ours $\StrengthFactor$ = 1.0 & $34.83 \pm 2.01$ & $0.91 \pm 0.03$ & $0.12 \pm 0.06$ \\
Ours $\StrengthFactor$ = 1.2 & $33.26 \pm 2.00$ & $0.87 \pm 0.03$ & $0.15 \pm 0.07$ \\
 \midrule
\bottomrule
\end{tabular}
\end{table}

\subsection{Robustness}

We evaluate the robustness of DeepSignature with respect to different transformations of signed images. We report the watermark verification rate WVR (see Section~\ref{sec:method:watermark_verification}), which includes error-correction, and is the practically relevant metric---measuring whether the watermark $\Watermark$ was successfully recovered and verified, but without guaranteeing image integrity, which must still be checked separately. Additionally, we report the bit error rate $BER$ of the watermark decoder as the ratio of incorrectly decoded bits from the payload $\PayloadLong$.

\begin{figure}[!t]
    \centering
    \includegraphics[width=1.2\colwidth]{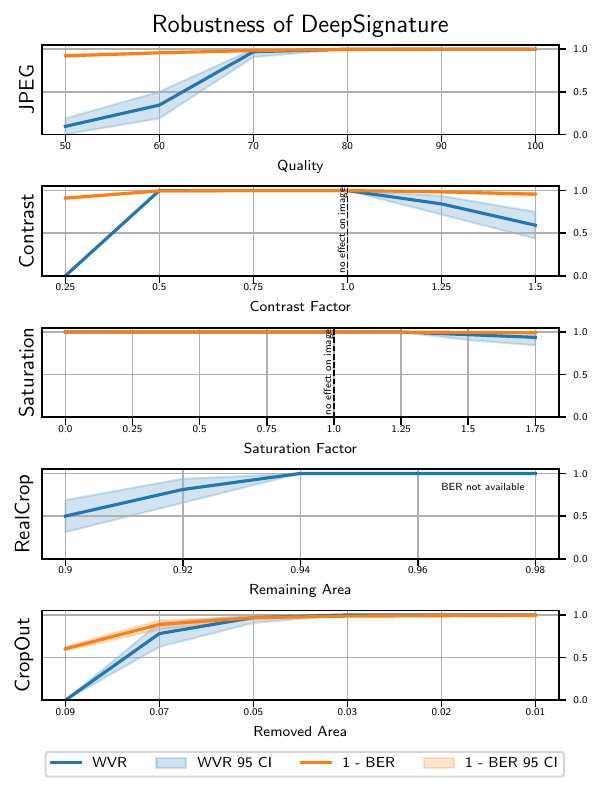}
    \caption{
    DeepSignature remains robust to various types of benign image transformations. The figure shows panels for different transformations, from top to bottom: JPEG compression, contrast changes, saturation adjustments, (random) cropping, and (random) crop outs (replacing pixels in a rectangular area). We measure (Y-axis) the bit error rate (BER, actually $1 - BER$), and the watermark verification rate (WVR), including 95\% confidence intervals (CI). The X-axis represents variations in the transformations. Saturation and contrast factors interpolate relative to the original image: a saturation factor of 0 yields a grayscale image, 1 leaves the image unchanged, and values greater than 1 increase saturation. A contrast factor of 0 yields a uniform gray image, 1 leaves the image unchanged, and values greater than 1 increase contrast.
    \berDisclaimer
    }
    \label{fig:robustness_clic}
\end{figure}

Figure~\ref{fig:robustness_clic} shows the method's robustness against various transformations. Both the watermark verification rate (WVR) and the bit error rate (BER) remain unaffected for JPEG transformations with quality factors of 80 or higher; however, these metrics begin to decline at a quality factor of 70 or below. In fact, only 41\% of the images processed at JPEG quality 60 had verifiable watermarks. DeepSignature remains robust under contrast changes, particularly when the contrast is lowered. Changes in saturation have little effect; remarkably, reducing saturation to the point where images become gray scale does not lower robustness. To assess robustness to cropping, we added 100 border pixels (see Appendix~\ref{sec:cropresist} for technical details) and observed high robustness when cropping 6\% of the total image area. Robustness to crop-outs (i.e., replacing pixels in rectangular areas) starts to drop when an area larger than 3\% is affected, at 9\%, no watermarks can be verified. Tables with results are in appendix \ref{sec:additional_results:robustness}.

To adjust the robustness of watermark recovery, strength factor $\StrengthFactor$ and scaling factor $\ScalingFactor$ can be adjusted. While $\StrengthFactor$ directly increases robustness by trading against imperceptibility, $\ScalingFactor$ enables more error-correction and trades increased robustness against lower integrity verification capability. We report additional results in appendix \ref{sec:additional_results:robustness}.

\begin{figure}
    \centering
    \includegraphics[width=1.4\colwidth]{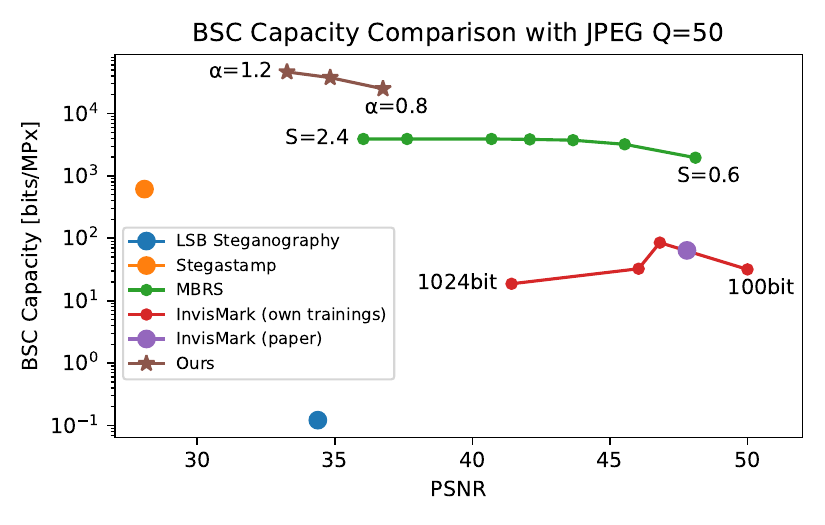}
    \caption{Under JPEG compression (quality 50), DeepSignature supports substantially higher payload capacity than the compared methods at comparable image quality. Figure compares capacity and image quality between different data hiding methods under JPEG compression with quality 50. All values were evaluated on the CLIC2024 test set. The capacity of DeepSignature is around 10 times larger than the MBRS checkpoint with comparable image quality. For InvisMark (paper) \citep{xuInvisMarkInvisibleRobust2024}, StegaStamp \citep{tancikStegaStampInvisibleHyperlinks2020} and MBRS \citep{jiaMBRSEnhancingRobustness2021} checkpoints provided by the respective authors were used. Additional models with different message lengths were trained for InvisMark (own trainings). Both MBRS and Deep Signature provide a strength factor to control the tradeoff between image fidelity and robustness. Note the extremely poor capacity of LSB steganography owed to the lack of robustness to JPEG compression.}
    \label{fig:hiding_comparison}
\end{figure}

\subsection{Sensitivity of Latent-Space Verification}

\begin{figure}[!t]
    \centering
    \includegraphics[width=1.2\colwidth]{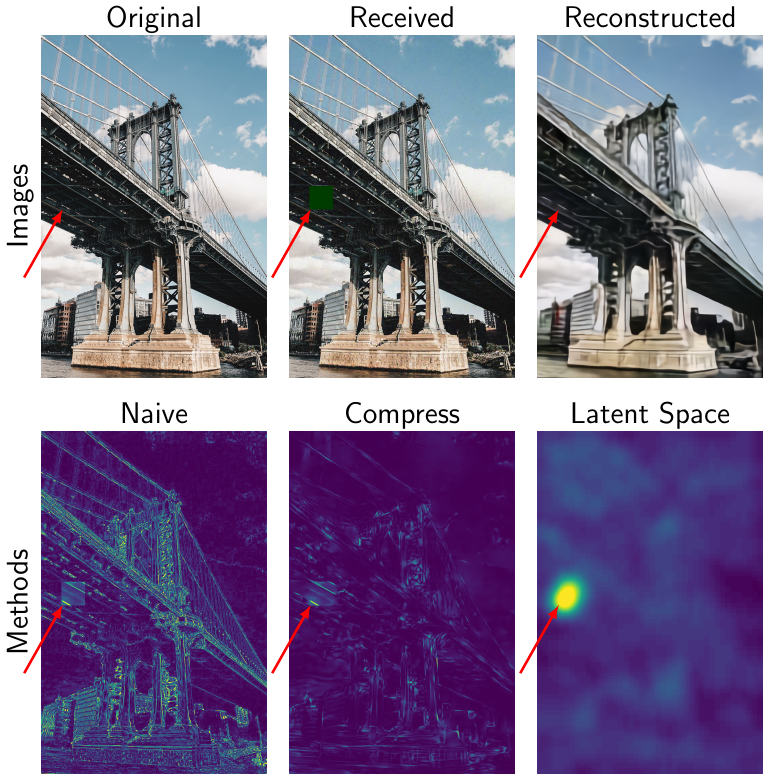}
    \caption{
    Latent-space verification accurately highlights tampered regions without being influenced by spurious artifacts. The figure compares different methods to verify image integrity using an example image that was tampered with. Top row, from left to right: the original image (Original), the received and tampered with image (Received) in which a small green rectangle was added (red arrow), and the reconstructed original image (Reconstructed). In the bottom row we compare the different methods by highlighting local differences. From left to right: comparing the reconstructed original image with the received image in pixel-space (Naive), comparing the reconstructed original image to the encoded and decoded received image (Compress), comparing both in latent-space by encoding the receive image (Latent Space). The results show that the manipulated region is barely visible (Naive), not the only region with irregularities (Compress), or clearly highlighted (Latent Space).
    Created using an image from CLIC 2022 licensed under the Unsplash License.
    }
    \label{fig:latent_verification_options}
\end{figure}

There are different options to verify the integrity of a signed image $\ReceivedImage$.
A naive approach would be to decode the latent representation of the original image to pixel-space: $\OriginalImage \approx \ReconstructedImage = \ContentDecoder(\ContentDecoderInput)$. The decoded image $\ReconstructedImage$ could then be compared to the received image $\ReceivedImage$ in pixel-space. This approach is sub-optimal because we compare a (strongly) compressed version $\ReconstructedImage$ with an un- or mildly compressed version $\ReceivedImage$. Our tests produced no satisfactory thresholds of image comparison metrics that lead to acceptable precision and recall trade-offs to separate between authentic and forged images.
Another option is to encode and decode the received image $\ContentDecoder(\ContentEncoder(\ReceivedImage))$, which introduces compression artifacts to the received image and ensures the comparison is conducted in the distribution modeled by the decoder. This improves comparison; however, we observed issues at locations with sharp edges. In our tests, comparing the latent representation of the original image $\ContentDecoderInput$ with the latent representation of the received (and potentially altered) image $\ContentDecoderInputReceived$ performed best. See Figure~\ref{fig:latent_verification_options} for a comparison of the different approaches using an example image. To check integrity in latent-space, we calculate change map $\ChangeMap$ and tampering score $\TamperingScore$ (see Section~\ref{sec:method:image_integrity}).

\begin{figure}[!t]
    \centering
    \includegraphics[width=1.4\colwidth]{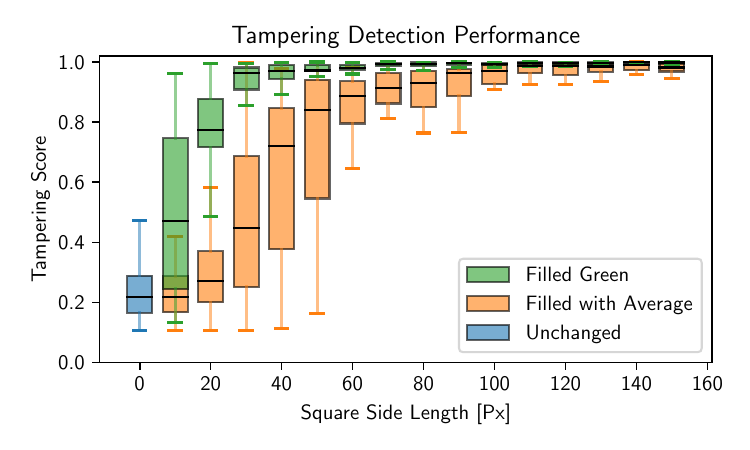}
    \caption{
    Tampering scores reliably detect tampering when manipulating square regions as small as 30 pixels side length. The figure shows how tampering scores change when manipulating images in square regions (adding rectangles of a
    specific side length) at random locations.
    Green boxes depict the result of changing each pixel inside the square to bright
    green, orange boxes the result of changing each pixel to the average of the
    image region inside the box, blue depicts the tampering scores of images without tampering (side length 0 pixels).
    Green squares are reliably detected at a smaller size because the visual
    difference is much greater.
    Images from the CLIC 2024 test set resized to $768 \times 768$.
    }
    \label{fig:latent_space_rectangles}
\end{figure}

We manipulate signed images by adding rectangles at random locations and of varying side length by either replacing the pixels of the manipulated area with i) green pixels or ii) the average pixel value of the manipulated region. Figure~\ref{fig:latent_space_rectangles} shows the results in terms of the tampering score $\TamperingScore$ on the CLIC 2024 test set, with all images resized to $768 \times 768$ pixels. Detecting green rectangles is easier and works for any manipulated area larger than $20 \times 20$ pixels using threshold $\TamperingScoreThreshold=0.7$. Using the average pixel value makes detection harder. Still, such manipulations are generally detected when they cover more than $60 \times 60$ pixels. Note the intentional use of the absolute size in pixels instead of stating manipulated area as a percentage of the image because the sensitivity is a function of the receptive field of the image encoder and decoder and therefore independent of the image size. Figure~\ref{fig:latent_space_rectangles} also shows tampering scores for images without any manipulation (side length of 0 pixels, marked blue). Surprisingly, these scores are not zero. This reflects the changes induced by the watermarking process ($\Delta = \OriginalImage - \WatermarkedImage$) and by the quantization operation, as we compare a quantized latent representation of the original image $\ContentDecoderInput$ with an non-quantized version of the received image $\ContentDecoderInputReceived$.

\subsection{Forgery Detection With Latent-Space Verification}

We evaluate DeepSignature on signed-image forgery detection and localization using CASIA V1.0, CASIA V2.0~\citep{dongCASIAImageTampering2013}, and Emu Edit~\citep{sheyninEmuEditPrecise2024}. Unlike blind forensic methods, DeepSignature operates with access to the original image content via the embedded watermark. For each authentic image $\mathbf{x}$ and corresponding forgery $\mathbf{x_f}$, we first sign $\mathbf{x}$ to obtain $\mathbf{x_w}$ and then create a tampered image $\mathbf{x_t}$ by replacing forged pixels in $\mathbf{x_w}$ according to a tampering mask $\mathbf{M} \in \{0, 1\}^{H \times W}$:
\begin{equation}
    \mathbf{x_t} = \mathbf{M} \odot \mathbf{x_f} + (1 - \mathbf{M}) \odot \mathbf{x_w} \, .
\end{equation}
For CASIA, $\mathbf{M}$ is given by the dataset. For Emu Edit, $\mathbf{M}$ is derived from thresholded pixel differences between $\mathbf{x}$ and $\mathbf{x_f}$, followed by removal of very small connected components. Consequently, CASIA and Emu Edit define localization targets differently: CASIA masks are operation-defined, whereas Emu Edit masks are difference-defined and often noisier.

Table~\ref{table:forgery_detection} summarizes the main results. Watermark verification succeeds for all authentic images, while only 35\%--51\% of forged images remain verifiable. The tampering score separates authentic from forged samples very well, yielding an AUC of 1.00 for both CASIA datasets and 0.98 for Emu Edit. Using an F1-optimized threshold, all samples in CASIA V1.0 are classified correctly, while false-positive / false-negative rates are 0.4\% / 0.7\% for CASIA V2.0 and 0.8\% / 5.4\% for Emu Edit. Tampering localization is strong on CASIA, with mean IoU of 0.64 for V1.0 and 0.61 for V2.0, but substantially lower on Emu Edit at 0.36.

\begin{table}[t]
  \caption{Forgery Detection: Shown are Watermark Verification Rates (WVR), Tampering Score (mean and  std), Tampering Localization with IoU (mean and std) for each authentic and forged images and different datasets. We also report the AUC when treating the identification of authentic / forged images as a binary classification score. We optimize F1-score and define the optimal threshold $\TamperingScoreThreshold$ for the classification and report the proportion of images verified with DeepSignature (DeepSignature Verified).}
  \label{table:forgery_detection}
  \centering
  \small
  \begin{tabular}{lcccccc}
    \toprule
    & \multicolumn{2}{c}{CASIA V1.0} & \multicolumn{2}{c}{CASIA V2.0} & \multicolumn{2}{c}{EMU Edit}  \\
    \cmidrule(lr){2-7}
    & Authentic & Forged & Authentic & Forged & Authentic & Forged \\
    \midrule
    Number of Images & 372 & 457 & 852 & 1619 & 1999 & 2022 \\
    Watermark Verification Rate & 100\% & 44\% & 100\% & 51\% & 100\% & 35\% \\
    Tampering Score & $0.28 \pm 0.09$& $0.98 \pm 0.02$ & $0.27 \pm 0.09$ & $0.97 \pm 0.06$ & $0.27 \pm 0.10$ & $0.93 \pm 0.16$ \\
    Tampering Localization (IoU) & & $0.64 \pm 0.10$ &  & $0.61 \pm 0.15$ & & $0.36 \pm 0.23$ \\
    AUC (on verified) & \multicolumn{2}{c}{1.00} & \multicolumn{2}{c}{1.00}  & \multicolumn{2}{c}{0.98}\\
    Optimal $\TamperingScoreThreshold$ for F1& \multicolumn{2}{c}{0.74} & \multicolumn{2}{c}{0.63}  & \multicolumn{2}{c}{0.56}\\
    \midrule
    \textbf{DeepSignature Verified} & 100\% & 0\% & 99.6\% & 0.7\% & 99.2\% & 5.4\% \\
    \bottomrule
  \end{tabular}
\end{table}

To better understand the Emu Edit gap, we further analyze subsets defined by edit type and edit strength. We exclude diffuse edit types (\textit{global}, \textit{style}) and use PSNR between original and edited images as a proxy for edit strength, where lower PSNR indicates stronger edits. In addition to IoU, we report any-overlap recall, which measures whether the predicted tampering region overlaps the target mask at all and is therefore less sensitive to noisy or fragmented masks. As shown in Table~\ref{table:emu_edit_localized}, restricting Emu Edit to localized edits of at least medium strength increases mean IoU from 0.36 to 0.47 and any-overlap recall from 0.90 to 1.00. This indicates that a substantial part of the aggregate Emu Edit gap is caused by weak or non-localizable edits rather than complete failure on meaningful local manipulations. Nevertheless, performance remains below CASIA, confirming that Emu Edit is intrinsically more challenging due to its noisier, difference-defined targets. Manual inspection of remaining false negatives showed that several low-score edits were barely perceptible by eye (see Appendix~\ref{appendix:forgery_localization}).

\begin{table}[t]
  \caption{Tampering localization performance. IoU and any-overlap recall across datasets and EMU Edit subsets. Excluding diffuse edit types (global, style) and restricting to medium/strong edits (based on PSNR) substantially improves EMU Edit localization, although performance remains below CASIA.}
  \label{table:emu_edit_localized}
  \centering
  \small
  \begin{tabular}{llcc}
    \toprule
    Subset & $N$ & IoU Mean$\uparrow$ & Any-Overlap$\uparrow$ \\
    \midrule
    CASIA V1.0 & 202 & 0.64 & 1.00 \\
    CASIA V2.0 & 830 & 0.61 & 0.98 \\
    \midrule
    EMU Edit (all) & 702 & 0.36 & 0.90 \\
    EMU Edit (localized + med./strong) & 440 & 0.47 & 1.00 \\
    \bottomrule
  \end{tabular}
\end{table}

Direct comparison to blind tampering-localization methods is limited. Recent methods such as TruFor~\citep{guillaroTruForLeveragingAllRound2023}, DiffForensics~\citep{yuDiffForensicsLeveragingDiffusion2024}, and UnionFormer~\citep{liUnionFormerUnifiedLearningTransformer2024} are evaluated in a blind-forensics setting and commonly report pixel-level AUC and/or pixel-level F1 for localization. By contrast, DeepSignature operates on signed images with access to the original through the embedded watermark, and our tampered test images are generated from signed authentic images using benchmark tampering masks rather than taken directly from the benchmark. Localization is evaluated here using IoU with respect to the reference tampering masks, the results should therefore be interpreted within the signed-image verification setting rather than as directly comparable blind-forensics benchmarks.

\subsection{Reconstruction Quality}

DeepSignature's reconstruction quality is primarily constrained by the available bit budget for the compact representation. While the hiding capacity of $\frac{1}{16}$ bit per pixel is remarkable for robust data hiding (see Fig.~\ref{fig:hiding_comparison}), only half of it is allocated to the compact representation $\BinaryEncoding$ (for $\ScalingFactor = 1$), the other half is used for metadata and channel coding to ensure robustness. As a result, reconstructions remain noticeably lossy. We quantify this effect in Table~\ref{table:reconstructionquality}, reporting PSNR/SSIM/LPIPS between $\OriginalImage$ and $\ReconstructedImage$ for different scaling factors $\ScalingFactor$.

\begin{table}[hbt]
  \caption{Image reconstruction quality measures. We report discrepancy between $\OriginalImage$ and $\ReconstructedImage$ using different metrics under variations of scaling factor $\ScalingFactor$. $\uparrow$ (higher is better) and $\downarrow$ (lower is better). Results are reported on the CLIC 2024 test set.}
  \label{table:reconstructionquality}
  \centering
  \small
  \setlength{\tabcolsep}{5pt}
\begin{tabular}{rrrr}
\toprule
 & PSNR$\uparrow$ & SSIM$\uparrow$ & LPIPS$\downarrow$ \\
 \midrule
     Ours $\ScalingFactor$=0.8 & $24.79 \pm 4.45$ & $0.73 \pm 0.18$ & $0.39 \pm 0.16$ \\
     Ours $\ScalingFactor$=1.0 & $25.50 \pm 4.42$ & $0.75 \pm 0.17$ & $0.35 \pm 0.14$ \\
     Ours $\ScalingFactor$=1.2 & $26.06 \pm 4.34$ & $0.76 \pm 0.16$ & $0.31 \pm 0.13$ \\
\midrule
\bottomrule
\end{tabular}
\end{table}

\subsection{Attack Scenarios}

We evaluate different attack scenarios by considering combinations of attacker goals and their level of access to DeepSignature. An attacker may have one or more of the following objectives:
\begin{enumerate}
    \item \textbf{Invalidate:} Render the signature invalid. The verification process must fail at the cryptographic signature verification stage.
    \item \textbf{Detect:} Determine if an image was signed.
    \item \textbf{Remove:} Eliminate all traces of the signature. Verification must fail and the output of the content decoder is indistinguishable from that of an unsigned image.
    \item \textbf{Extract:} Extract the hidden watermark from a signed image.
    \item \textbf{Forge:} Generate a valid DeepSignature without access to the private key.
    \item \textbf{Collision:} Modify a signed image while preserving its signature, potentially via watermark extraction and re-embedding. This remains theoretically possible but was unsuccessful in our experiments.
\end{enumerate}

We classify attackers based on their level of access to DeepSignature and signed images:

\begin{enumerate}
    \item \textbf{No Access}: Access to a limited number of signed images.
    \item \textbf{Limited API Access}: Can create a few signed images per day via a rate-limited API.
    \item \textbf{Unlimited API Access}: Can generate large numbers of signed images but lacks source code.
    \item \textbf{Source Access}: Has source code but not model weights.
    \item \textbf{Full Access}: Possesses source code and model weights but not the private key.
\end{enumerate}

The feasibility of attacks depends on the chosen deployment model of DeepSignature. In a \textit{DeepSignature-as-a-Service} setup, where signers and verifiers interact via an API without access to the underlying model, attackers must reverse-engineer the watermark structure from a limited number of signed images. In contrast, a \textit{fully open-source deployment}, where both the source code and model weights are public, allows independent verification but also grants attackers full access, making certain attacks (e.g., extraction) more viable.

Our evaluation (see Table~\ref{table:attacks}) shows that invalidating a signature is consistently easy, while forging a signature remains impossible due to cryptographic guarantees. Detection and removal are generally feasible, particularly with increased access to signed images. Extracting the watermark is difficult with only source code access but becomes more viable when model weights are available. Collision attacks remain challenging, even with full access.
However, collision attacks could succeed by exploiting the inherent ambiguity in how similarity is defined and measured, particularly when benign transformations are allowed.

\begin{table}[t]
  \caption{Likelihood of Successful Attacks by Attacker Access Level.
  \textit{I} indicates attacks requiring cryptographic breaking, which is infeasible.}
  \label{table:attacks}
  \centering
  \small
  \renewcommand{\arraystretch}{1.1}
  \setlength{\tabcolsep}{4pt}
  \begin{tabular}{llccccc}
    \toprule
    \multirow{2}{*}{\textbf{}} & \multicolumn{5}{c}{\textbf{Access Level}} \\
    \cmidrule(lr){2-6}
    & \textbf{No} & \textbf{Limited} & \textbf{Unlimited} & \textbf{Source} & \textbf{Full} \\
    \midrule
    \textbf{Invalidate} & H & H & H & H & H \\
    \textbf{Detect}     & M & H & H & H & H \\
    \textbf{Remove}     & L & L & M & M & H \\
    \textbf{Extract}    & L & L & L & M & H \\
    \textbf{Collision}  & L & L & L & L & M \\
    \textbf{Forge}      & I & I & I & I & I \\
    \bottomrule
  \end{tabular}

  \vspace{2mm}
  {\textbf{Likelihood:} H = High, M = Medium, L = Low, I = Impossible.}
\end{table}

\subsection{Adjustability}\label{sec:adjustaility}

DeepSignature is designed with modular components, making it adaptable to different application-specific requirements. Replacing the content autoencoder can optimize compression ratio and manipulation sensitivity, while retraining the watermark encoders and decoders enhances imperceptibility, ensuring adaptability across use cases and deployment scenarios. The approach, as outlined in Figure~\ref{fig:teaser}, generalizes to other modalities, such as audio or video, and can embed additional metadata, e.g., trusted timestamps or content-specific markers.

\subsection{Limitations}
\label{sec:limitations}

Depending on application-specific requirements, DeepSignature might not be able to cover a specified trade-off between capacity, robustness, and imperceptibility. Particularly, if there are strict requirements regarding imperceptibility, robustness might be low, or the fidelity encoded by the latent representation might not satisfy the desired protection level.

There is no inherent guarantee that a specific image $\OriginalImageSpecific$ signed with DeepSignature withstands a set of transformations deemed benign. It might become necessary to test the robustness of a specific sample and increase robustness by adjusting method-specific parameters for sample $\OriginalImageSpecific$. Adjusting DeepSignature based on individual samples requires additional resources and complicates the approach.

Image integrity verification in latent space inevitably leaves some room for tampering---within bounds imposed by the content encoder. Exactly what these bounds are and what guarantees they might provide requires further study.

Additionally, the effectiveness of DeepSignature is influenced by image resolution. The watermark consists of both fixed-size and content-specific components, and smaller images provide less capacity for encoding. Due to this limitation, DeepSignature is better suited for higher-resolution images and may not be applicable to images below a certain size\footnote{The technical limit is $233 \times 233$ pixels}.

\section{Conclusion}

\noindent We present DeepSignature, a novel approach to sign images for authenticity verification. By embedding digitally signed content-encoding watermarks into images using neural networks, DeepSignature enables reliable verification even after common transformations such as lossy compression. Notably, it operates without introducing a new image format, making it compatible with existing standards and workflows.

DeepSignature's robustness ensures that images can be transmitted over various communication channels without requiring special handling. The signature remains intact even when images are re-encoded or stripped of metadata, preserving the signature despite common transformations. Importantly, intermediaries do not need to take any special measures or even be aware of the signature for it to remain intact.

Tampering attempts are effectively detected. Significant modifications destroy the content-encoding watermark, causing verification failure, while cryptographic guarantees prevent manipulation of the watermark itself. Minor edits allow for error-free extraction, enabling direct comparison between the embedded content and the received image. Our results demonstrate that forgery attempts can be accurately detected and localized.

\section*{Acknowledgments}
This work was supported in part by the Swiss National Science Foundation (grant number 190708) and in part by InnoSuisse (project number 100.698 IP-ICT).

\appendix

\section{Additional Results}
\label{sec:additional_results}
\label{sec:additional_results:robustness}

Table~\ref{table:robustness_factors} shows the results of different combinations for strength factor and scaling factor. Particularly, the strength factor has a significant effect, as seen for JPEG 70 and WVR.

\begin{table}[!t]
  \caption{Robustness to transformations under variations of strength factor $\StrengthFactor$ and scaling factor $\ScalingFactor$. Shown are bit error rate (BER) and watermark verification rate (WVR). Calculated on the CLIC 2024 test set.}
  \label{table:robustness_factors}
  \centering
  \small
  \setlength{\tabcolsep}{5pt}
\begin{tabular}{llllll}
\toprule
 &  & \multicolumn{2}{c}{BER} & \multicolumn{2}{c}{WVR} \\
 &  & JPEG 70 & JPEG 80 & JPEG 70 & JPEG 80 \\
$s$ & $\alpha$ &  &  &  &  \\
\midrule
\multirow[t]{3}{*}{0.6} & 0.8 & $0.05 \pm 0.02$ & $0.02 \pm 0.01$ & $0.72 \pm 0.46$ & $1.00 \pm 0.00$ \\
 & 1.0 & $0.02 \pm 0.01$ & $0.00 \pm 0.00$ & $1.00 \pm 0.00$ & $1.00 \pm 0.00$ \\
 & 1.2 & $0.01 \pm 0.01$ & $0.00 \pm 0.00$ & $1.00 \pm 0.00$ & $1.00 \pm 0.00$ \\
\cline{1-6}
\multirow[t]{3}{*}{0.8} & 0.8 & $0.05 \pm 0.02$ & $0.02 \pm 0.01$ & $0.47 \pm 0.51$ & $1.00 \pm 0.00$ \\
 & 1.0 & $0.02 \pm 0.01$ & $0.00 \pm 0.00$ & $1.00 \pm 0.00$ & $1.00 \pm 0.00$ \\
 & 1.2 & $0.01 \pm 0.00$ & $0.00 \pm 0.00$ & $1.00 \pm 0.00$ & $1.00 \pm 0.00$ \\
\cline{1-6}
\multirow[t]{3}{*}{1.0} & 0.8 & $0.05 \pm 0.02$ & $0.02 \pm 0.01$ & $0.28 \pm 0.46$ & $1.00 \pm 0.00$ \\
 & 1.0 & $0.02 \pm 0.01$ & $0.00 \pm 0.00$ & $0.97 \pm 0.18$ & $1.00 \pm 0.00$ \\
 & 1.2 & $0.01 \pm 0.00$ & $0.00 \pm 0.00$ & $1.00 \pm 0.00$ & $1.00 \pm 0.00$ \\
\bottomrule
\end{tabular}
\end{table}

Table \ref{table:robustness_transformations_coco} shows robustness to different transformations on a sample of the COCO 2017 test set (1,000 images).

\begin{table}[!t]
  \caption{Robustness to transformations. We use scaling factor $\ScalingFactor=1.0$ and strength factor $\StrengthFactor=1$. Shown are bit error rate (BER) and watermark verification rate (WVR) for different transformations and their transformation-specific strength. Calculated on 1,000 samples of the COCO 2017 test set, resizing images to $768 \times 768$ pixels. We add 50 border pixels for the experiments with RealCrop and PseudoCrop. (*)\berDisclaimer}
  \label{table:robustness_transformations_coco}
  \centering
  \small
  \setlength{\tabcolsep}{5pt}
\begin{tabular}{llll}
\toprule
\multicolumn{4}{c}{\textbf{COCO 2017}}\\
\midrule
 &  & BER & WVR \\
Noise & Noise Strength &  &  \\
\midrule
\multirow[t]{9}{*}{Contrast} & 0.25 & $0.09 \pm 0.03$ & $0.00 \pm 0.07$ \\
 & 0.5 & $0.00 \pm 0.00$ & $1.00 \pm 0.00$ \\
 & 0.75 & $0.00 \pm 0.00$ & $1.00 \pm 0.00$ \\
 & 1.0 & $0.00 \pm 0.00$ & $1.00 \pm 0.00$ \\
 & 1.25 & $0.03 \pm 0.04$ & $0.66 \pm 0.47$ \\
 & 1.5 & $0.06 \pm 0.06$ & $0.33 \pm 0.47$ \\
 & 1.75 & $0.09 \pm 0.07$ & $0.18 \pm 0.39$ \\
\cline{1-4}
\multirow[t]{6}{*}{CropOut} & 0.01 & $0.01 \pm 0.04$ & $0.99 \pm 0.09$ \\
 & 0.02 & $0.04 \pm 0.10$ & $0.94 \pm 0.24$ \\
 & 0.03 & $0.04 \pm 0.10$ & $0.94 \pm 0.24$ \\
 & 0.05 & $0.06 \pm 0.11$ & $0.92 \pm 0.28$ \\
 & 0.07 & $0.24 \pm 0.21$ & $0.49 \pm 0.50$ \\
 & 0.09 & $0.43 \pm 0.08$ & $0.00 \pm 0.00$ \\
\cline{1-4}
\multirow[t]{6}{*}{JPEG} & 50 & $0.05 \pm 0.03$ & $0.37 \pm 0.48$ \\
 & 60 & $0.02 \pm 0.02$ & $0.72 \pm 0.45$ \\
 & 70 & $0.01 \pm 0.01$ & $0.99 \pm 0.08$ \\
 & 80 & $0.00 \pm 0.00$ & $1.00 \pm 0.00$ \\
 & 90 & $0.00 \pm 0.00$ & $1.00 \pm 0.00$ \\
 & 100 & $0.00 \pm 0.00$ & $1.00 \pm 0.00$ \\
\cline{1-4}
\multirow[t]{5}{*}{PseudoCrop} & 0.9 & $0.07 \pm 0.09$ & $0.93 \pm 0.26$ \\
 & 0.92 & $0.04 \pm 0.04$ & $0.99 \pm 0.11$ \\
 & 0.94 & $0.03 \pm 0.00$ & $1.00 \pm 0.00$ \\
 & 0.96 & $0.02 \pm 0.00$ & $1.00 \pm 0.00$ \\
 & 0.98 & $0.01 \pm 0.00$ & $1.00 \pm 0.00$ \\
\cline{1-4}
\multirow[t]{5}{*}{RealCrop} & 0.9 & * & $0.76 \pm 0.43$ \\
 & 0.92 & * & $0.94 \pm 0.23$ \\
 & 0.94 & * & $1.00 \pm 0.00$ \\
 & 0.98 & * & $1.00 \pm 0.00$ \\
\cline{1-4}
\multirow[t]{8}{*}{Saturation}
& 0.0 & $0.00 \pm 0.00$ & $1.00 \pm 0.00$ \\
 & 0.5 & $0.00 \pm 0.00$ & $1.00 \pm 0.00$ \\
 & 0.75 & $0.00 \pm 0.00$ & $1.00 \pm 0.00$ \\
 & 1.0 & $0.00 \pm 0.00$ & $1.00 \pm 0.00$ \\
 & 1.25 & $0.00 \pm 0.00$ & $1.00 \pm 0.00$ \\
 & 1.75 & $0.00 \pm 0.01$ & $0.99 \pm 0.08$ \\
\bottomrule
\end{tabular}
\end{table}

\section{Implementation Details}
\label{sec:implementation_details}

\subsection{Payload Composition}
\label{sec:payload_composition}

\begin{figure}[!t]
    \centering
    \includegraphics[width=1.4\colwidth]{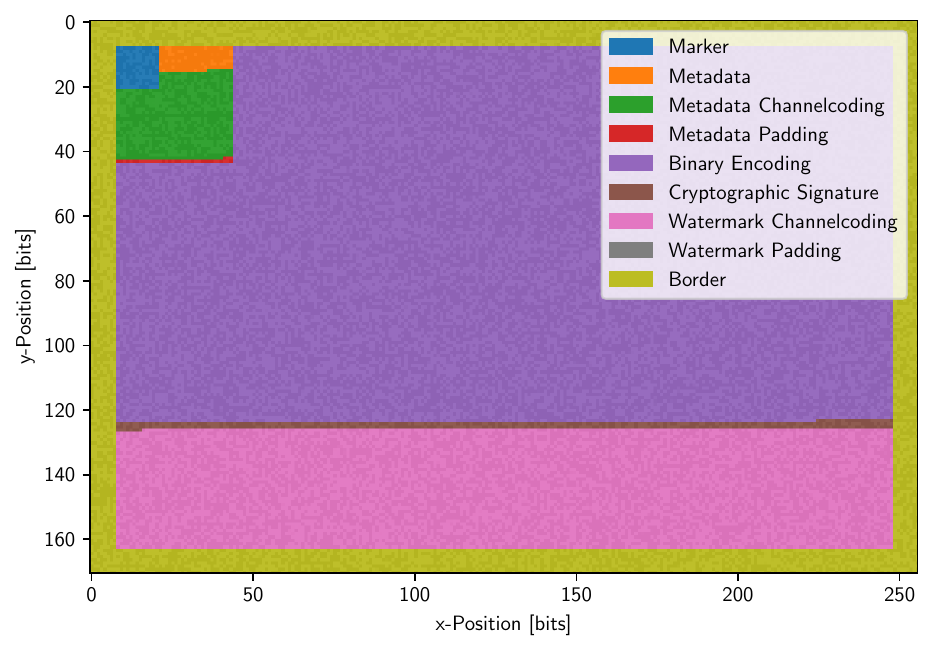}
    \caption{Payload composition for an example image of size $1024
    \times 768$ with a border of 32 pixels. The payload is of size $1024 / 4 \times 768 / 4$. The marker is used to find the
    start of the metadata block. The metadata block is encoded with fixed
    channel coding parameters and contains information about the message stride
    and length of the rest of the message. Compact representation and channel
    coding take up the majority of the capacity.}
    \label{fig:hidden-message}
\end{figure}

The payload $\PayloadLong$ is reshaped to a 2D binary tensor of shape
$\frac{H}{2^n} \times \frac{W}{2^n} \, ,$
where $n$ is the number of upscaling blocks in the watermark encoder
$\WatermarkEncoder$.
This tensor comprises the binary encoding $\BinaryEncoding$, the
signature $\Signature$, the metadata $\Metadata$, and channel coding information.
If necessary, random padding bits are added.
Figure~\ref{fig:hidden-message} illustrates the payload's composition.

The metadata $\Metadata$ encapsulates image-specific parameters for channel coding and facilitates the implementation of both the scaling factor $\ScalingFactor$ and crop resistance.
It includes the following elements: channel coding parameters $\mathbf{c}$, the length of the channel coded watermark $\ErrorCorrection_{\mathbf{w}}(\Watermark)$, the width $\ScalingFactor \cdot \mathbf{W'}$ and height $\ScalingFactor \cdot \mathbf{H'}$ of the reconstructed image $\ReconstructedImage$, the stride and height of the watermark $\Watermark$ (excluding its border), and the width of the border on each side (see Appendix~\ref{sec:cropresist}).

\subsection{Watermark Decoding and Crop Resistance} \label{sec:cropresist}

Cropping, whether benign or malicious, presents challenges for DeepSignature, primarily due to information loss in cropped regions. One solution is reserving a border where no information is embedded, however, mixing regions with and without embedded information might introduce local artifacts (sudden spatial quality changes). To mitigate this, random bits are embedded in the border area (see Section~\nameref{sec:payload_composition}).

To detect cropping, a marker is added to the payload. It must be large enough to minimize false positives but not so large that it reduces payload capacity substantially, particularly relevant for smaller images. We use a $13~\times~13$ bits marker detected via cross-correlation.

The watermark decoder $\WatermarkDecoder$ employs strided convolutions to extract the watermark, but these are not shift-equivariant for non-multiples of the stride. If cropping misaligns the payload, decoding fails. With a global stride of 4, there are $4 \times 4 = 16$ possible offsets, hence we use an exhaustive search to recover the payload.

\subsection{Forgery Localization} \label{appendix:forgery_localization}

\begin{figure}[!t]
    \centering
    \includegraphics[width=1.4\colwidth]{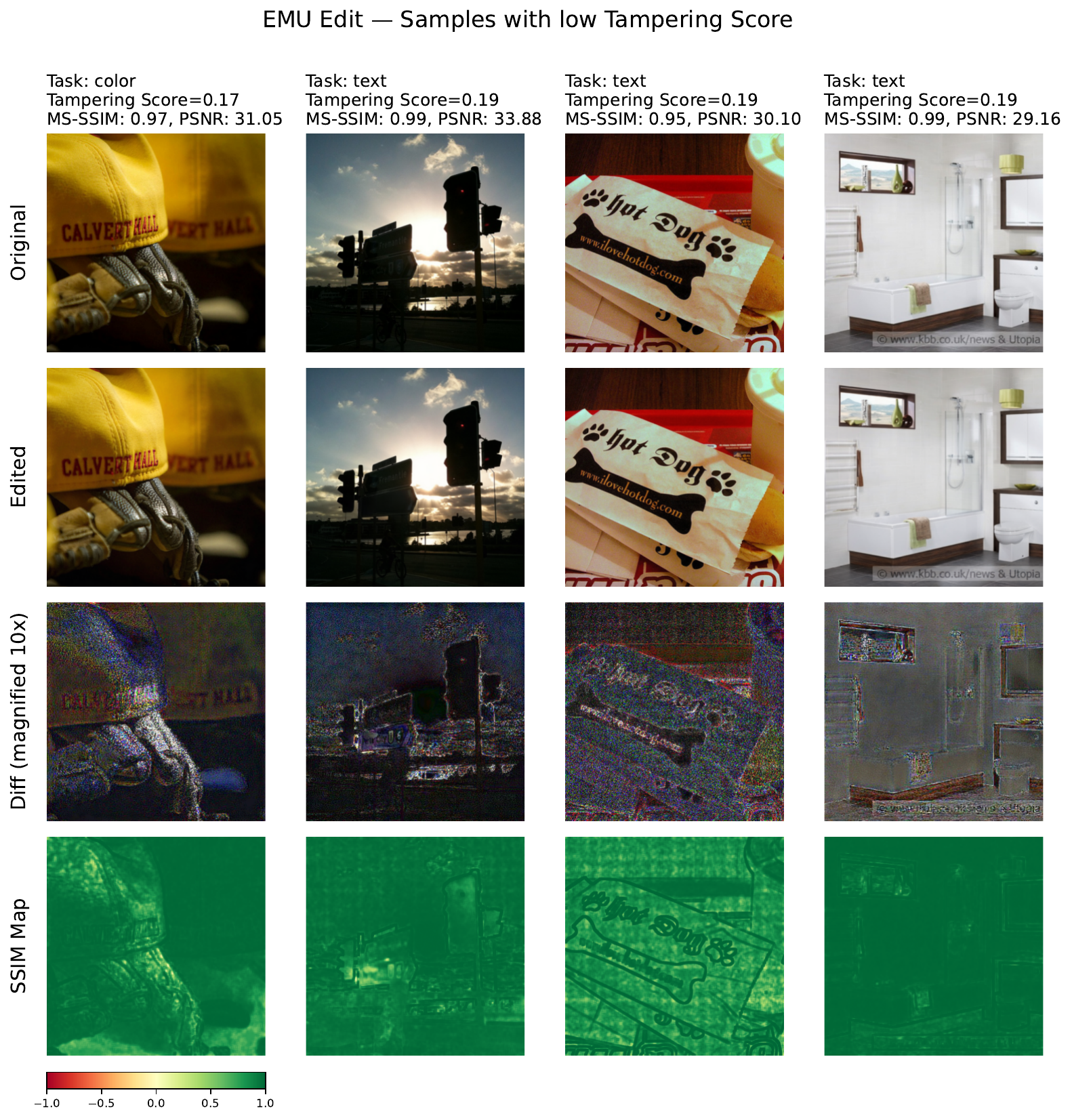}
    \caption{Some Emu Edit forgeries are barely perceptible. Shown are example image pairs from Emu Edit that were manipulated only very subtly. Such cases can appear as false negatives, although the edits are hardly visible even under close inspection. From top to bottom: the original image, the edited image, the magnified absolute difference between original and edited, and the local SSIM map between original and edited.}
    \label{fig:emu_edit_low_tampering_samples}
\end{figure}

Figure \ref{fig:emu_edit_low_tampering_samples} shows sample images from Emu Edit for which DeepSignature verification was successful even though these were edited. Visual inspection shows no apparent difference between original and edited versions and thus are in fact correctly classified as true negatives.

\clearpage

\bibliographystyle{unsrtnat}
\bibliography{references}

\end{document}

%% file: notation.tex
\newcommand{\OriginalImage}{\mathbf{x}}
\newcommand{\OriginalImageLong}{\OriginalImage \in \mathbb{R}^{3 \times H \times W}}
\newcommand{\OriginalImageSpecific}{\OriginalImage^{(i)}}

\newcommand{\WatermarkedImage}{\mathbf{x_w}}

\newcommand{\ReceivedImage}{\OriginalImage'}

\newcommand{\ReconstructedImage}{\mathbf{\hat{x}}}

\newcommand{\CodebookDimension}{D}
\newcommand{\ContentEncoderOutput}{\mathbf{z_e}}
\newcommand{\ContentEncoderOutputLong}{\ContentEncoderOutput \in \mathbb{R}^{\CodebookDimension \times H' \times W'}}

\newcommand{\QuantizedRepresentation}{\mathbf{z_i}}
\newcommand{\QuantizedRepresentationLong}{\QuantizedRepresentation \in \mathbb{N}^{H' \times W'}}

\newcommand{\ContentDecoderInput}{\mathbf{z_q}}
\newcommand{\ContentDecoderInputLong}{\ContentDecoderInput \in \mathbb{R}^{\CodebookDimension \times H' \times W'}}

\newcommand{\ContentDecoderInputReceived}{\ContentEncoderOutput'}

\newcommand{\BinaryEncoding}{\mathbf{z_{b}}}
\newcommand{\BinaryEncodingLong}{\BinaryEncoding \in \{0, 1\}^l}

\newcommand{\Signature}{\mathbf{s_b}}

\newcommand{\Watermark}{\mathbf{w_b}}
\newcommand{\WatermarkLong}{\Watermark \in \{0, 1\}^m}

\newcommand{\Payload}{\mathbf{w_p}}
\newcommand{\PayloadLong}{\Payload \in \{0, 1\}^n}

\newcommand{\Metadata}{\mathbf{w_m}}
\newcommand{\MetadataLong}{\Metadata \in \{0, 1\}^k}

\newcommand{\DecodedWatermark}{\mathbf{\hat{w}_b}}
\newcommand{\DecodedWatermarkLongBeforeQuant}{\DecodedWatermark \in \mathbb{R}^{m}}  

\newcommand{\ContentEncoder}{E_{c,\theta}}

\newcommand{\ContentDecoder}{D_{c,\phi}}

\newcommand{\Quantizer}{Q}

\newcommand{\CodebookSize}{K}

\newcommand{\Codebook}{\mathbf{C}}
\newcommand{\CodebookLong}{\Codebook \in \mathbb{R}^{\CodebookSize \times \CodebookDimension}}

\newcommand{\WatermarkEncoder}{E_{w,\theta}}

\newcommand{\WatermarkDecoder}{D_{w,\phi}}

\newcommand{\AdversarialNetwork}{A_{\theta}}

\newcommand{\StrengthFactor}{\alpha}

\newcommand{\ScalingFactor}{s}
\newcommand{\ScalingFactorLong}{\ScalingFactor \in \mathbb{R}}

\newcommand{\NoiseFunction}{N}

\newcommand{\DownsamplingFactor}{S}

\newcommand{\PrivateKey}{\mathbf{k_s}}
\newcommand{\PublicKey}{\mathbf{k_p}}
\newcommand{\ErrorCorrection}{\xi}
\newcommand{\ErrorCorrectionDecoding}{\ErrorCorrection^{-1}}
\newcommand{\ErrorCorrectionMetadata}{\ErrorCorrection_{\mathbf{m}}}

\newcommand{\ErrorCorrectionWatermark}{\ErrorCorrection_{\mathbf{w}}}

\newcommand{\TamperingScore}{TS}
\newcommand{\TamperingScoreLong}{\TamperingScore \in [0, 1]}

\newcommand{\TamperingScoreLocal}{TS_{\text{local}}}
\newcommand{\TamperingScoreGlobal}{TS_{\text{global}}}
\newcommand{\TamperingScoreWeight}{\beta_g}

\newcommand{\TamperingScoreThreshold}{\tau_{ts}}

\newcommand{\ChangeMap}{\mathbf{z_{cm}}}
\newcommand{\ChangeMapLong}{\ChangeMap \in \mathbb{R}^{1 \times H' \times W'}}
\newcommand{\ChangeMapAverage}{\mathbf{\bar{z}_{cm}}}

\newcommand{\concat}{~\|~}